\documentclass[12pt,a4paper,superscriptaddress]{revtex4}
\usepackage{graphicx}
\usepackage{graphics}
\usepackage{amsmath}
\usepackage{dcolumn}
\usepackage{amssymb}
\usepackage{bm}
\usepackage[latin1]{inputenc}
\newcommand{\be}{\begin{equation}}
\newcommand{\ee}{\end{equation}}
\newcommand{\bea}{\begin{eqnarray}}
\newcommand{\eea}{\end{eqnarray}}

\newcommand{\pa}{\partial}

\begin{document}
\immediate\write16{<<WARNING: LINEDRAW macros work with emTeX-dvivers
                    and other drivers supporting emTeX \special's
                    (dviscr, dvihplj, dvidot, dvips, dviwin, etc.) >>}

%% Macros for drawing Feynman graphs and other complex diagrams
%% Designed by A.V.Voronin (1993); modified in 1995
%% Steklov Math. Inst., e-mail: av@voronin.mian.su
%%
\newdimen\Lengthunit       \Lengthunit  = 1.5cm
\newcount\Nhalfperiods     \Nhalfperiods= 9
\newcount\magnitude        \magnitude = 1000

\catcode`\*=11
\newdimen\L*   \newdimen\d*   \newdimen\d**
\newdimen\dm*  \newdimen\dd*  \newdimen\dt*
\newdimen\a*   \newdimen\b*   \newdimen\c*
\newdimen\a**  \newdimen\b**
\newdimen\xL*  \newdimen\yL*
\newdimen\rx*  \newdimen\ry*
\newdimen\tmp* \newdimen\linwid*

\newcount\k*   \newcount\l*   \newcount\m*
\newcount\k**  \newcount\l**  \newcount\m**
\newcount\n*   \newcount\dn*  \newcount\r*
\newcount\N*   \newcount\*one \newcount\*two  \*one=1 \*two=2
\newcount\*ths \*ths=1000
\newcount\angle*  \newcount\q*  \newcount\q**
\newcount\angle** \angle**=0
\newcount\sc*     \sc*=0

\newtoks\cos*  \cos*={1}
\newtoks\sin*  \sin*={0}

\catcode`\[=13

\def\rotate(#1){\advance\angle**#1\angle*=\angle**
\q**=\angle*\ifnum\q**<0\q**=-\q**\fi
\ifnum\q**>360\q*=\angle*\divide\q*360\multiply\q*360\advance\angle*-\q*\fi
\ifnum\angle*<0\advance\angle*360\fi\q**=\angle*\divide\q**90\q**=\q**
\def\sgcos*{+}\def\sgsin*{+}\relax
\ifcase\q**\or
 \def\sgcos*{-}\def\sgsin*{+}\or
 \def\sgcos*{-}\def\sgsin*{-}\or
 \def\sgcos*{+}\def\sgsin*{-}\else\fi
\q*=\q**
\multiply\q*90\advance\angle*-\q*
\ifnum\angle*>45\sc*=1\angle*=-\angle*\advance\angle*90\else\sc*=0\fi
\def[##1,##2]{\ifnum\sc*=0\relax
\edef\cs*{\sgcos*.##1}\edef\sn*{\sgsin*.##2}\ifcase\q**\or
 \edef\cs*{\sgcos*.##2}\edef\sn*{\sgsin*.##1}\or
 \edef\cs*{\sgcos*.##1}\edef\sn*{\sgsin*.##2}\or
 \edef\cs*{\sgcos*.##2}\edef\sn*{\sgsin*.##1}\else\fi\else
\edef\cs*{\sgcos*.##2}\edef\sn*{\sgsin*.##1}\ifcase\q**\or
 \edef\cs*{\sgcos*.##1}\edef\sn*{\sgsin*.##2}\or
 \edef\cs*{\sgcos*.##2}\edef\sn*{\sgsin*.##1}\or
 \edef\cs*{\sgcos*.##1}\edef\sn*{\sgsin*.##2}\else\fi\fi
\cos*={\cs*}\sin*={\sn*}\global\edef\gcos*{\cs*}\global\edef\gsin*{\sn*}}\relax
\ifcase\angle*[9999,0]\or
[999,017]\or[999,034]\or[998,052]\or[997,069]\or[996,087]\or
[994,104]\or[992,121]\or[990,139]\or[987,156]\or[984,173]\or
[981,190]\or[978,207]\or[974,224]\or[970,241]\or[965,258]\or
[961,275]\or[956,292]\or[951,309]\or[945,325]\or[939,342]\or
[933,358]\or[927,374]\or[920,390]\or[913,406]\or[906,422]\or
[898,438]\or[891,453]\or[882,469]\or[874,484]\or[866,499]\or
[857,515]\or[848,529]\or[838,544]\or[829,559]\or[819,573]\or
[809,587]\or[798,601]\or[788,615]\or[777,629]\or[766,642]\or
[754,656]\or[743,669]\or[731,681]\or[719,694]\or[707,707]\or
\else[9999,0]\fi}

\catcode`\[=12

\def\GRAPH(hsize=#1)#2{\hbox to #1\Lengthunit{#2\hss}}

\def\Linewidth#1{\global\linwid*=#1\relax
\global\divide\linwid*10\global\multiply\linwid*\mag
\global\divide\linwid*100\special{em:linewidth \the\linwid*}}

\Linewidth{.4pt}
\def\sm*{\special{em:moveto}}
\def\sl*{\special{em:lineto}}
\let\moveto=\sm*
\let\lineto=\sl*
\newbox\spm*   \newbox\spl*
\setbox\spm*\hbox{\sm*}
\setbox\spl*\hbox{\sl*}

\def\mov#1(#2,#3)#4{\rlap{\L*=#1\Lengthunit
\xL*=#2\L* \yL*=#3\L*
\xL*=\xscale\xL* \yL*=\yscale\yL*
\rx* \the\cos*\xL* \tmp* \the\sin*\yL* \advance\rx*-\tmp*
\ry* \the\cos*\yL* \tmp* \the\sin*\xL* \advance\ry*\tmp*
\kern\rx*\raise\ry*\hbox{#4}}}

\def\rmov*(#1,#2)#3{\rlap{\xL*=#1\yL*=#2\relax
\rx* \the\cos*\xL* \tmp* \the\sin*\yL* \advance\rx*-\tmp*
\ry* \the\cos*\yL* \tmp* \the\sin*\xL* \advance\ry*\tmp*
\kern\rx*\raise\ry*\hbox{#3}}}

\def\lin#1(#2,#3){\rlap{\sm*\mov#1(#2,#3){\sl*}}}

\def\arr*(#1,#2,#3){\rmov*(#1\dd*,#1\dt*){\sm*
\rmov*(#2\dd*,#2\dt*){\rmov*(#3\dt*,-#3\dd*){\sl*}}\sm*
\rmov*(#2\dd*,#2\dt*){\rmov*(-#3\dt*,#3\dd*){\sl*}}}}

\def\arrow#1(#2,#3){\rlap{\lin#1(#2,#3)\mov#1(#2,#3){\relax
\d**=-.012\Lengthunit\dd*=#2\d**\dt*=#3\d**
\arr*(1,10,4)\arr*(3,8,4)\arr*(4.8,4.2,3)}}}

\def\arrlin#1(#2,#3){\rlap{\L*=#1\Lengthunit\L*=.5\L*
\lin#1(#2,#3)\rmov*(#2\L*,#3\L*){\arrow.1(#2,#3)}}}

\def\dasharrow#1(#2,#3){\rlap{{\Lengthunit=0.9\Lengthunit
\dashlin#1(#2,#3)\mov#1(#2,#3){\sm*}}\mov#1(#2,#3){\sl*
\d**=-.012\Lengthunit\dd*=#2\d**\dt*=#3\d**
\arr*(1,10,4)\arr*(3,8,4)\arr*(4.8,4.2,3)}}}

\def\clap#1{\hbox to 0pt{\hss #1\hss}}

\def\ind(#1,#2)#3{\rlap{\L*=.1\Lengthunit
\xL*=#1\L* \yL*=#2\L*
\rx* \the\cos*\xL* \tmp* \the\sin*\yL* \advance\rx*-\tmp*
\ry* \the\cos*\yL* \tmp* \the\sin*\xL* \advance\ry*\tmp*
\kern\rx*\raise\ry*\hbox{\lower2pt\clap{$#3$}}}}

\def\sh*(#1,#2)#3{\rlap{\dm*=\the\n*\d**
\xL*=\xscale\dm* \yL*=\yscale\dm* \xL*=#1\xL* \yL*=#2\yL*
\rx* \the\cos*\xL* \tmp* \the\sin*\yL* \advance\rx*-\tmp*
\ry* \the\cos*\yL* \tmp* \the\sin*\xL* \advance\ry*\tmp*
\kern\rx*\raise\ry*\hbox{#3}}}

\def\calcnum*#1(#2,#3){\a*=1000sp\b*=1000sp\a*=#2\a*\b*=#3\b*
\ifdim\a*<0pt\a*-\a*\fi\ifdim\b*<0pt\b*-\b*\fi
\ifdim\a*>\b*\c*=.96\a*\advance\c*.4\b*
\else\c*=.96\b*\advance\c*.4\a*\fi
\k*\a*\multiply\k*\k*\l*\b*\multiply\l*\l*
\m*\k*\advance\m*\l*\n*\c*\r*\n*\multiply\n*\n*
\dn*\m*\advance\dn*-\n*\divide\dn*2\divide\dn*\r*
\advance\r*\dn*
\c*=\the\Nhalfperiods5sp\c*=#1\c*\ifdim\c*<0pt\c*-\c*\fi
\multiply\c*\r*\N*\c*\divide\N*10000}

\def\dashlin#1(#2,#3){\rlap{\calcnum*#1(#2,#3)\relax
\d**=#1\Lengthunit\ifdim\d**<0pt\d**-\d**\fi
\divide\N*2\multiply\N*2\advance\N*\*one
\divide\d**\N*\sm*\n*\*one\sh*(#2,#3){\sl*}\loop
\advance\n*\*one\sh*(#2,#3){\sm*}\advance\n*\*one
\sh*(#2,#3){\sl*}\ifnum\n*<\N*\repeat}}

\def\dashdotlin#1(#2,#3){\rlap{\calcnum*#1(#2,#3)\relax
\d**=#1\Lengthunit\ifdim\d**<0pt\d**-\d**\fi
\divide\N*2\multiply\N*2\advance\N*1\multiply\N*2\relax
\divide\d**\N*\sm*\n*\*two\sh*(#2,#3){\sl*}\loop
\advance\n*\*one\sh*(#2,#3){\kern-1.48pt\lower.5pt\hbox{\rm.}}\relax
\advance\n*\*one\sh*(#2,#3){\sm*}\advance\n*\*two
\sh*(#2,#3){\sl*}\ifnum\n*<\N*\repeat}}

\def\shl*(#1,#2)#3{\kern#1#3\lower#2#3\hbox{\unhcopy\spl*}}

\def\trianglin#1(#2,#3){\rlap{\toks0={#2}\toks1={#3}\calcnum*#1(#2,#3)\relax
\dd*=.57\Lengthunit\dd*=#1\dd*\divide\dd*\N*
\divide\dd*\*ths \multiply\dd*\magnitude
\d**=#1\Lengthunit\ifdim\d**<0pt\d**-\d**\fi
\multiply\N*2\divide\d**\N*\sm*\n*\*one\loop
\shl**{\dd*}\dd*-\dd*\advance\n*2\relax
\ifnum\n*<\N*\repeat\n*\N*\shl**{0pt}}}

\def\wavelin#1(#2,#3){\rlap{\toks0={#2}\toks1={#3}\calcnum*#1(#2,#3)\relax
\dd*=.23\Lengthunit\dd*=#1\dd*\divide\dd*\N*
\divide\dd*\*ths \multiply\dd*\magnitude
\d**=#1\Lengthunit\ifdim\d**<0pt\d**-\d**\fi
\multiply\N*4\divide\d**\N*\sm*\n*\*one\loop
\shl**{\dd*}\dt*=1.3\dd*\advance\n*\*one
\shl**{\dt*}\advance\n*\*one
\shl**{\dd*}\advance\n*\*two
\dd*-\dd*\ifnum\n*<\N*\repeat\n*\N*\shl**{0pt}}}

\def\w*lin(#1,#2){\rlap{\toks0={#1}\toks1={#2}\d**=\Lengthunit\dd*=-.12\d**
\divide\dd*\*ths \multiply\dd*\magnitude
\N*8\divide\d**\N*\sm*\n*\*one\loop
\shl**{\dd*}\dt*=1.3\dd*\advance\n*\*one
\shl**{\dt*}\advance\n*\*one
\shl**{\dd*}\advance\n*\*one
\shl**{0pt}\dd*-\dd*\advance\n*1\ifnum\n*<\N*\repeat}}

\def\l*arc(#1,#2)[#3][#4]{\rlap{\toks0={#1}\toks1={#2}\d**=\Lengthunit
\dd*=#3.037\d**\dd*=#4\dd*\dt*=#3.049\d**\dt*=#4\dt*\ifdim\d**>10mm\relax
\d**=.25\d**\n*\*one\shl**{-\dd*}\n*\*two\shl**{-\dt*}\n*3\relax
\shl**{-\dd*}\n*4\relax\shl**{0pt}\else
\ifdim\d**>5mm\d**=.5\d**\n*\*one\shl**{-\dt*}\n*\*two
\shl**{0pt}\else\n*\*one\shl**{0pt}\fi\fi}}

\def\d*arc(#1,#2)[#3][#4]{\rlap{\toks0={#1}\toks1={#2}\d**=\Lengthunit
\dd*=#3.037\d**\dd*=#4\dd*\d**=.25\d**\sm*\n*\*one\shl**{-\dd*}\relax
\n*3\relax\sh*(#1,#2){\xL*=\xscale\dd*\yL*=\yscale\dd*
\kern#2\xL*\lower#1\yL*\hbox{\sm*}}\n*4\relax\shl**{0pt}}}

\def\shl**#1{\c*=\the\n*\d**\d*=#1\relax
\a*=\the\toks0\c*\b*=\the\toks1\d*\advance\a*-\b*
\b*=\the\toks1\c*\d*=\the\toks0\d*\advance\b*\d*
\a*=\xscale\a*\b*=\yscale\b*
\rx* \the\cos*\a* \tmp* \the\sin*\b* \advance\rx*-\tmp*
\ry* \the\cos*\b* \tmp* \the\sin*\a* \advance\ry*\tmp*
\raise\ry*\rlap{\kern\rx*\unhcopy\spl*}}

\def\wlin*#1(#2,#3)[#4]{\rlap{\toks0={#2}\toks1={#3}\relax
\c*=#1\l*\c*\c*=.01\Lengthunit\m*\c*\divide\l*\m*
\c*=\the\Nhalfperiods5sp\multiply\c*\l*\N*\c*\divide\N*\*ths
\divide\N*2\multiply\N*2\advance\N*\*one
\dd*=.002\Lengthunit\dd*=#4\dd*\multiply\dd*\l*\divide\dd*\N*
\divide\dd*\*ths \multiply\dd*\magnitude
\d**=#1\multiply\N*4\divide\d**\N*\sm*\n*\*one\loop
\shl**{\dd*}\dt*=1.3\dd*\advance\n*\*one
\shl**{\dt*}\advance\n*\*one
\shl**{\dd*}\advance\n*\*two
\dd*-\dd*\ifnum\n*<\N*\repeat\n*\N*\shl**{0pt}}}

\def\wavebox#1{\setbox0\hbox{#1}\relax
\a*=\wd0\advance\a*14pt\b*=\ht0\advance\b*\dp0\advance\b*14pt\relax
\hbox{\kern9pt\relax
\rmov*(0pt,\ht0){\rmov*(-7pt,7pt){\wlin*\a*(1,0)[+]\wlin*\b*(0,-1)[-]}}\relax
\rmov*(\wd0,-\dp0){\rmov*(7pt,-7pt){\wlin*\a*(-1,0)[+]\wlin*\b*(0,1)[-]}}\relax
\box0\kern9pt}}

\def\rectangle#1(#2,#3){\relax
\lin#1(#2,0)\lin#1(0,#3)\mov#1(0,#3){\lin#1(#2,0)}\mov#1(#2,0){\lin#1(0,#3)}}

\def\dashrectangle#1(#2,#3){\dashlin#1(#2,0)\dashlin#1(0,#3)\relax
\mov#1(0,#3){\dashlin#1(#2,0)}\mov#1(#2,0){\dashlin#1(0,#3)}}

\def\waverectangle#1(#2,#3){\L*=#1\Lengthunit\a*=#2\L*\b*=#3\L*
\ifdim\a*<0pt\a*-\a*\def\x*{-1}\else\def\x*{1}\fi
\ifdim\b*<0pt\b*-\b*\def\y*{-1}\else\def\y*{1}\fi
\wlin*\a*(\x*,0)[-]\wlin*\b*(0,\y*)[+]\relax
\mov#1(0,#3){\wlin*\a*(\x*,0)[+]}\mov#1(#2,0){\wlin*\b*(0,\y*)[-]}}

\def\calcparab*{\ifnum\n*>\m*\k*\N*\advance\k*-\n*\else\k*\n*\fi
\a*=\the\k* sp\a*=10\a*\b*\dm*\advance\b*-\a*\k*\b*
\a*=\the\*ths\b*\divide\a*\l*\multiply\a*\k*
\divide\a*\l*\k*\*ths\r*\a*\advance\k*-\r*\dt*=\the\k*\L*}

\def\arcto#1(#2,#3)[#4]{\rlap{\toks0={#2}\toks1={#3}\calcnum*#1(#2,#3)\relax
\dm*=135sp\dm*=#1\dm*\d**=#1\Lengthunit\ifdim\dm*<0pt\dm*-\dm*\fi
\multiply\dm*\r*\a*=.3\dm*\a*=#4\a*\ifdim\a*<0pt\a*-\a*\fi
\advance\dm*\a*\N*\dm*\divide\N*10000\relax
\divide\N*2\multiply\N*2\advance\N*\*one
\L*=-.25\d**\L*=#4\L*\divide\d**\N*\divide\L*\*ths
\m*\N*\divide\m*2\dm*=\the\m*5sp\l*\dm*\sm*\n*\*one\loop
\calcparab*\shl**{-\dt*}\advance\n*1\ifnum\n*<\N*\repeat}}

\def\arrarcto#1(#2,#3)[#4]{\L*=#1\Lengthunit\L*=.54\L*
\arcto#1(#2,#3)[#4]\rmov*(#2\L*,#3\L*){\d*=.457\L*\d*=#4\d*\d**-\d*
\rmov*(#3\d**,#2\d*){\arrow.02(#2,#3)}}}

\def\dasharcto#1(#2,#3)[#4]{\rlap{\toks0={#2}\toks1={#3}\relax
\calcnum*#1(#2,#3)\dm*=\the\N*5sp\a*=.3\dm*\a*=#4\a*\ifdim\a*<0pt\a*-\a*\fi
\advance\dm*\a*\N*\dm*
\divide\N*20\multiply\N*2\advance\N*1\d**=#1\Lengthunit
\L*=-.25\d**\L*=#4\L*\divide\d**\N*\divide\L*\*ths
\m*\N*\divide\m*2\dm*=\the\m*5sp\l*\dm*
\sm*\n*\*one\loop\calcparab*
\shl**{-\dt*}\advance\n*1\ifnum\n*>\N*\else\calcparab*
\sh*(#2,#3){\xL*=#3\dt* \yL*=#2\dt*
\rx* \the\cos*\xL* \tmp* \the\sin*\yL* \advance\rx*\tmp*
\ry* \the\cos*\yL* \tmp* \the\sin*\xL* \advance\ry*-\tmp*
\kern\rx*\lower\ry*\hbox{\sm*}}\fi
\advance\n*1\ifnum\n*<\N*\repeat}}

\def\*shl*#1{\c*=\the\n*\d**\advance\c*#1\a**\d*\dt*\advance\d*#1\b**
\a*=\the\toks0\c*\b*=\the\toks1\d*\advance\a*-\b*
\b*=\the\toks1\c*\d*=\the\toks0\d*\advance\b*\d*
\rx* \the\cos*\a* \tmp* \the\sin*\b* \advance\rx*-\tmp*
\ry* \the\cos*\b* \tmp* \the\sin*\a* \advance\ry*\tmp*
\raise\ry*\rlap{\kern\rx*\unhcopy\spl*}}

\def\calcnormal*#1{\b**=10000sp\a**\b**\k*\n*\advance\k*-\m*
\multiply\a**\k*\divide\a**\m*\a**=#1\a**\ifdim\a**<0pt\a**-\a**\fi
\ifdim\a**>\b**\d*=.96\a**\advance\d*.4\b**
\else\d*=.96\b**\advance\d*.4\a**\fi
\d*=.01\d*\r*\d*\divide\a**\r*\divide\b**\r*
\ifnum\k*<0\a**-\a**\fi\d*=#1\d*\ifdim\d*<0pt\b**-\b**\fi
\k*\a**\a**=\the\k*\dd*\k*\b**\b**=\the\k*\dd*}

\def\wavearcto#1(#2,#3)[#4]{\rlap{\toks0={#2}\toks1={#3}\relax
\calcnum*#1(#2,#3)\c*=\the\N*5sp\a*=.4\c*\a*=#4\a*\ifdim\a*<0pt\a*-\a*\fi
\advance\c*\a*\N*\c*\divide\N*20\multiply\N*2\advance\N*-1\multiply\N*4\relax
\d**=#1\Lengthunit\dd*=.012\d**
\divide\dd*\*ths \multiply\dd*\magnitude
\ifdim\d**<0pt\d**-\d**\fi\L*=.25\d**
\divide\d**\N*\divide\dd*\N*\L*=#4\L*\divide\L*\*ths
\m*\N*\divide\m*2\dm*=\the\m*0sp\l*\dm*
\sm*\n*\*one\loop\calcnormal*{#4}\calcparab*
\*shl*{1}\advance\n*\*one\calcparab*
\*shl*{1.3}\advance\n*\*one\calcparab*
\*shl*{1}\advance\n*2\dd*-\dd*\ifnum\n*<\N*\repeat\n*\N*\shl**{0pt}}}

\def\triangarcto#1(#2,#3)[#4]{\rlap{\toks0={#2}\toks1={#3}\relax
\calcnum*#1(#2,#3)\c*=\the\N*5sp\a*=.4\c*\a*=#4\a*\ifdim\a*<0pt\a*-\a*\fi
\advance\c*\a*\N*\c*\divide\N*20\multiply\N*2\advance\N*-1\multiply\N*2\relax
\d**=#1\Lengthunit\dd*=.012\d**
\divide\dd*\*ths \multiply\dd*\magnitude
\ifdim\d**<0pt\d**-\d**\fi\L*=.25\d**
\divide\d**\N*\divide\dd*\N*\L*=#4\L*\divide\L*\*ths
\m*\N*\divide\m*2\dm*=\the\m*0sp\l*\dm*
\sm*\n*\*one\loop\calcnormal*{#4}\calcparab*
\*shl*{1}\advance\n*2\dd*-\dd*\ifnum\n*<\N*\repeat\n*\N*\shl**{0pt}}}

\def\hr*#1{\L*=\xscale\Lengthunit\ifnum
\angle**=0\clap{\vrule width#1\L* height.1pt}\else
\L*=#1\L*\L*=.5\L*\rmov*(-\L*,0pt){\sm*}\rmov*(\L*,0pt){\sl*}\fi}

\def\shade#1[#2]{\rlap{\Lengthunit=#1\Lengthunit
\special{em:linewidth .001pt}\relax
\mov(0,#2.05){\hr*{.994}}\mov(0,#2.1){\hr*{.980}}\relax
\mov(0,#2.15){\hr*{.953}}\mov(0,#2.2){\hr*{.916}}\relax
\mov(0,#2.25){\hr*{.867}}\mov(0,#2.3){\hr*{.798}}\relax
\mov(0,#2.35){\hr*{.715}}\mov(0,#2.4){\hr*{.603}}\relax
\mov(0,#2.45){\hr*{.435}}\special{em:linewidth \the\linwid*}}}

\def\dshade#1[#2]{\rlap{\special{em:linewidth .001pt}\relax
\Lengthunit=#1\Lengthunit\if#2-\def\t*{+}\else\def\t*{-}\fi
\mov(0,\t*.025){\relax
\mov(0,#2.05){\hr*{.995}}\mov(0,#2.1){\hr*{.988}}\relax
\mov(0,#2.15){\hr*{.969}}\mov(0,#2.2){\hr*{.937}}\relax
\mov(0,#2.25){\hr*{.893}}\mov(0,#2.3){\hr*{.836}}\relax
\mov(0,#2.35){\hr*{.760}}\mov(0,#2.4){\hr*{.662}}\relax
\mov(0,#2.45){\hr*{.531}}\mov(0,#2.5){\hr*{.320}}\relax
\special{em:linewidth \the\linwid*}}}}

\def\vdot{\rlap{\kern-1.9pt\lower1.8pt\hbox{$\scriptstyle\bullet$}}}
\def\vtimes{\rlap{\kern-3pt\lower1.8pt\hbox{$\scriptstyle\times$}}}
\def\vDot{\rlap{\kern-2.3pt\lower2.7pt\hbox{$\bullet$}}}
\def\vTimes{\rlap{\kern-3.6pt\lower2.4pt\hbox{$\times$}}}

\def\arc(#1)[#2,#3]{{\k*=#2\l*=#3\m*=\l*
\advance\m*-6\ifnum\k*>\l*\relax\else
{\rotate(#2)\mov(#1,0){\sm*}}\loop
\ifnum\k*<\m*\advance\k*5{\rotate(\k*)\mov(#1,0){\sl*}}\repeat
{\rotate(#3)\mov(#1,0){\sl*}}\fi}}

\def\dasharc(#1)[#2,#3]{{\k**=#2\n*=#3\advance\n*-1\advance\n*-\k**
\L*=1000sp\L*#1\L* \multiply\L*\n* \multiply\L*\Nhalfperiods
\divide\L*57\N*\L* \divide\N*2000\ifnum\N*=0\N*1\fi
\r*\n*  \divide\r*\N* \ifnum\r*<2\r*2\fi
\m**\r* \divide\m**2 \l**\r* \advance\l**-\m** \N*\n* \divide\N*\r*
\k**\r* \multiply\k**\N* \dn*\n*
\advance\dn*-\k** \divide\dn*2\advance\dn*\*one
\r*\l** \divide\r*2\advance\dn*\r* \advance\N*-2\k**#2\relax
\ifnum\l**<6{\rotate(#2)\mov(#1,0){\sm*}}\advance\k**\dn*
{\rotate(\k**)\mov(#1,0){\sl*}}\advance\k**\m**
{\rotate(\k**)\mov(#1,0){\sm*}}\loop
\advance\k**\l**{\rotate(\k**)\mov(#1,0){\sl*}}\advance\k**\m**
{\rotate(\k**)\mov(#1,0){\sm*}}\advance\N*-1\ifnum\N*>0\repeat
{\rotate(#3)\mov(#1,0){\sl*}}\else\advance\k**\dn*
\arc(#1)[#2,\k**]\loop\advance\k**\m** \r*\k**
\advance\k**\l** {\arc(#1)[\r*,\k**]}\relax
\advance\N*-1\ifnum\N*>0\repeat
\advance\k**\m**\arc(#1)[\k**,#3]\fi}}

\def\triangarc#1(#2)[#3,#4]{{\k**=#3\n*=#4\advance\n*-\k**
\L*=1000sp\L*#2\L* \multiply\L*\n* \multiply\L*\Nhalfperiods
\divide\L*57\N*\L* \divide\N*1000\ifnum\N*=0\N*1\fi
\d**=#2\Lengthunit \d*\d** \divide\d*57\multiply\d*\n*
\r*\n*  \divide\r*\N* \ifnum\r*<2\r*2\fi
\m**\r* \divide\m**2 \l**\r* \advance\l**-\m** \N*\n* \divide\N*\r*
\dt*\d* \divide\dt*\N* \dt*.5\dt* \dt*#1\dt*
\divide\dt*1000\multiply\dt*\magnitude
\k**\r* \multiply\k**\N* \dn*\n* \advance\dn*-\k** \divide\dn*2\relax
\r*\l** \divide\r*2\advance\dn*\r* \advance\N*-1\k**#3\relax
{\rotate(#3)\mov(#2,0){\sm*}}\advance\k**\dn*
{\rotate(\k**)\mov(#2,0){\sl*}}\advance\k**-\m**\advance\l**\m**\loop\dt*-\dt*
\d*\d** \advance\d*\dt*
\advance\k**\l**{\rotate(\k**)\rmov*(\d*,0pt){\sl*}}%
\advance\N*-1\ifnum\N*>0\repeat\advance\k**\m**
{\rotate(\k**)\mov(#2,0){\sl*}}{\rotate(#4)\mov(#2,0){\sl*}}}}

\def\wavearc#1(#2)[#3,#4]{{\k**=#3\n*=#4\advance\n*-\k**
\L*=4000sp\L*#2\L* \multiply\L*\n* \multiply\L*\Nhalfperiods
\divide\L*57\N*\L* \divide\N*1000\ifnum\N*=0\N*1\fi
\d**=#2\Lengthunit \d*\d** \divide\d*57\multiply\d*\n*
\r*\n*  \divide\r*\N* \ifnum\r*=0\r*1\fi
\m**\r* \divide\m**2 \l**\r* \advance\l**-\m** \N*\n* \divide\N*\r*
\dt*\d* \divide\dt*\N* \dt*.7\dt* \dt*#1\dt*
\divide\dt*1000\multiply\dt*\magnitude
\k**\r* \multiply\k**\N* \dn*\n* \advance\dn*-\k** \divide\dn*2\relax
\divide\N*4\advance\N*-1\k**#3\relax
{\rotate(#3)\mov(#2,0){\sm*}}\advance\k**\dn*
{\rotate(\k**)\mov(#2,0){\sl*}}\advance\k**-\m**\advance\l**\m**\loop\dt*-\dt*
\d*\d** \advance\d*\dt* \dd*\d** \advance\dd*1.3\dt*
\advance\k**\r*{\rotate(\k**)\rmov*(\d*,0pt){\sl*}}\relax
\advance\k**\r*{\rotate(\k**)\rmov*(\dd*,0pt){\sl*}}\relax
\advance\k**\r*{\rotate(\k**)\rmov*(\d*,0pt){\sl*}}\relax
\advance\k**\r*
\advance\N*-1\ifnum\N*>0\repeat\advance\k**\m**
{\rotate(\k**)\mov(#2,0){\sl*}}{\rotate(#4)\mov(#2,0){\sl*}}}}

\def\gmov*#1(#2,#3)#4{\rlap{\L*=#1\Lengthunit
\xL*=#2\L* \yL*=#3\L*
\rx* \gcos*\xL* \tmp* \gsin*\yL* \advance\rx*-\tmp*
\ry* \gcos*\yL* \tmp* \gsin*\xL* \advance\ry*\tmp*
\rx*=\xscale\rx* \ry*=\yscale\ry*
\xL* \the\cos*\rx* \tmp* \the\sin*\ry* \advance\xL*-\tmp*
\yL* \the\cos*\ry* \tmp* \the\sin*\rx* \advance\yL*\tmp*
\kern\xL*\raise\yL*\hbox{#4}}}

\def\rgmov*(#1,#2)#3{\rlap{\xL*#1\yL*#2\relax
\rx* \gcos*\xL* \tmp* \gsin*\yL* \advance\rx*-\tmp*
\ry* \gcos*\yL* \tmp* \gsin*\xL* \advance\ry*\tmp*
\rx*=\xscale\rx* \ry*=\yscale\ry*
\xL* \the\cos*\rx* \tmp* \the\sin*\ry* \advance\xL*-\tmp*
\yL* \the\cos*\ry* \tmp* \the\sin*\rx* \advance\yL*\tmp*
\kern\xL*\raise\yL*\hbox{#3}}}

\def\Earc(#1)[#2,#3][#4,#5]{{\k*=#2\l*=#3\m*=\l*
\advance\m*-6\ifnum\k*>\l*\relax\else\def\xscale{#4}\def\yscale{#5}\relax
{\angle**0\rotate(#2)}\gmov*(#1,0){\sm*}\loop
\ifnum\k*<\m*\advance\k*5\relax
{\angle**0\rotate(\k*)}\gmov*(#1,0){\sl*}\repeat
{\angle**0\rotate(#3)}\gmov*(#1,0){\sl*}\relax
\def\xscale{1}\def\yscale{1}\fi}}

\def\dashEarc(#1)[#2,#3][#4,#5]{{\k**=#2\n*=#3\advance\n*-1\advance\n*-\k**
\L*=1000sp\L*#1\L* \multiply\L*\n* \multiply\L*\Nhalfperiods
\divide\L*57\N*\L* \divide\N*2000\ifnum\N*=0\N*1\fi
\r*\n*  \divide\r*\N* \ifnum\r*<2\r*2\fi
\m**\r* \divide\m**2 \l**\r* \advance\l**-\m** \N*\n* \divide\N*\r*
\k**\r*\multiply\k**\N* \dn*\n* \advance\dn*-\k** \divide\dn*2\advance\dn*\*one
\r*\l** \divide\r*2\advance\dn*\r* \advance\N*-2\k**#2\relax
\ifnum\l**<6\def\xscale{#4}\def\yscale{#5}\relax
{\angle**0\rotate(#2)}\gmov*(#1,0){\sm*}\advance\k**\dn*
{\angle**0\rotate(\k**)}\gmov*(#1,0){\sl*}\advance\k**\m**
{\angle**0\rotate(\k**)}\gmov*(#1,0){\sm*}\loop
\advance\k**\l**{\angle**0\rotate(\k**)}\gmov*(#1,0){\sl*}\advance\k**\m**
{\angle**0\rotate(\k**)}\gmov*(#1,0){\sm*}\advance\N*-1\ifnum\N*>0\repeat
{\angle**0\rotate(#3)}\gmov*(#1,0){\sl*}\def\xscale{1}\def\yscale{1}\else
\advance\k**\dn* \Earc(#1)[#2,\k**][#4,#5]\loop\advance\k**\m** \r*\k**
\advance\k**\l** {\Earc(#1)[\r*,\k**][#4,#5]}\relax
\advance\N*-1\ifnum\N*>0\repeat
\advance\k**\m**\Earc(#1)[\k**,#3][#4,#5]\fi}}

\def\triangEarc#1(#2)[#3,#4][#5,#6]{{\k**=#3\n*=#4\advance\n*-\k**
\L*=1000sp\L*#2\L* \multiply\L*\n* \multiply\L*\Nhalfperiods
\divide\L*57\N*\L* \divide\N*1000\ifnum\N*=0\N*1\fi
\d**=#2\Lengthunit \d*\d** \divide\d*57\multiply\d*\n*
\r*\n*  \divide\r*\N* \ifnum\r*<2\r*2\fi
\m**\r* \divide\m**2 \l**\r* \advance\l**-\m** \N*\n* \divide\N*\r*
\dt*\d* \divide\dt*\N* \dt*.5\dt* \dt*#1\dt*
\divide\dt*1000\multiply\dt*\magnitude
\k**\r* \multiply\k**\N* \dn*\n* \advance\dn*-\k** \divide\dn*2\relax
\r*\l** \divide\r*2\advance\dn*\r* \advance\N*-1\k**#3\relax
\def\xscale{#5}\def\yscale{#6}\relax
{\angle**0\rotate(#3)}\gmov*(#2,0){\sm*}\advance\k**\dn*
{\angle**0\rotate(\k**)}\gmov*(#2,0){\sl*}\advance\k**-\m**
\advance\l**\m**\loop\dt*-\dt* \d*\d** \advance\d*\dt*
\advance\k**\l**{\angle**0\rotate(\k**)}\rgmov*(\d*,0pt){\sl*}\relax
\advance\N*-1\ifnum\N*>0\repeat\advance\k**\m**
{\angle**0\rotate(\k**)}\gmov*(#2,0){\sl*}\relax
{\angle**0\rotate(#4)}\gmov*(#2,0){\sl*}\def\xscale{1}\def\yscale{1}}}

\def\waveEarc#1(#2)[#3,#4][#5,#6]{{\k**=#3\n*=#4\advance\n*-\k**
\L*=4000sp\L*#2\L* \multiply\L*\n* \multiply\L*\Nhalfperiods
\divide\L*57\N*\L* \divide\N*1000\ifnum\N*=0\N*1\fi
\d**=#2\Lengthunit \d*\d** \divide\d*57\multiply\d*\n*
\r*\n*  \divide\r*\N* \ifnum\r*=0\r*1\fi
\m**\r* \divide\m**2 \l**\r* \advance\l**-\m** \N*\n* \divide\N*\r*
\dt*\d* \divide\dt*\N* \dt*.7\dt* \dt*#1\dt*
\divide\dt*1000\multiply\dt*\magnitude
\k**\r* \multiply\k**\N* \dn*\n* \advance\dn*-\k** \divide\dn*2\relax
\divide\N*4\advance\N*-1\k**#3\def\xscale{#5}\def\yscale{#6}\relax
{\angle**0\rotate(#3)}\gmov*(#2,0){\sm*}\advance\k**\dn*
{\angle**0\rotate(\k**)}\gmov*(#2,0){\sl*}\advance\k**-\m**
\advance\l**\m**\loop\dt*-\dt*
\d*\d** \advance\d*\dt* \dd*\d** \advance\dd*1.3\dt*
\advance\k**\r*{\angle**0\rotate(\k**)}\rgmov*(\d*,0pt){\sl*}\relax
\advance\k**\r*{\angle**0\rotate(\k**)}\rgmov*(\dd*,0pt){\sl*}\relax
\advance\k**\r*{\angle**0\rotate(\k**)}\rgmov*(\d*,0pt){\sl*}\relax
\advance\k**\r*
\advance\N*-1\ifnum\N*>0\repeat\advance\k**\m**
{\angle**0\rotate(\k**)}\gmov*(#2,0){\sl*}\relax
{\angle**0\rotate(#4)}\gmov*(#2,0){\sl*}\def\xscale{1}\def\yscale{1}}}

\newcount\CatcodeOfAtSign
\CatcodeOfAtSign=\the\catcode`\@
\catcode`\@=11
\def\@arc#1[#2][#3]{\rlap{\Lengthunit=#1\Lengthunit
\sm*\l*arc(#2.1914,#3.0381)[#2][#3]\relax
\mov(#2.1914,#3.0381){\l*arc(#2.1622,#3.1084)[#2][#3]}\relax
\mov(#2.3536,#3.1465){\l*arc(#2.1084,#3.1622)[#2][#3]}\relax
\mov(#2.4619,#3.3086){\l*arc(#2.0381,#3.1914)[#2][#3]}}}

\def\dash@arc#1[#2][#3]{\rlap{\Lengthunit=#1\Lengthunit
\d*arc(#2.1914,#3.0381)[#2][#3]\relax
\mov(#2.1914,#3.0381){\d*arc(#2.1622,#3.1084)[#2][#3]}\relax
\mov(#2.3536,#3.1465){\d*arc(#2.1084,#3.1622)[#2][#3]}\relax
\mov(#2.4619,#3.3086){\d*arc(#2.0381,#3.1914)[#2][#3]}}}

\def\wave@arc#1[#2][#3]{\rlap{\Lengthunit=#1\Lengthunit
\w*lin(#2.1914,#3.0381)\relax
\mov(#2.1914,#3.0381){\w*lin(#2.1622,#3.1084)}\relax
\mov(#2.3536,#3.1465){\w*lin(#2.1084,#3.1622)}\relax
\mov(#2.4619,#3.3086){\w*lin(#2.0381,#3.1914)}}}

\def\bezier#1(#2,#3)(#4,#5)(#6,#7){\N*#1\l*\N* \advance\l*\*one
\d* #4\Lengthunit \advance\d* -#2\Lengthunit \multiply\d* \*two
\b* #6\Lengthunit \advance\b* -#2\Lengthunit
\advance\b*-\d* \divide\b*\N*
\d** #5\Lengthunit \advance\d** -#3\Lengthunit \multiply\d** \*two
\b** #7\Lengthunit \advance\b** -#3\Lengthunit
\advance\b** -\d** \divide\b**\N*
\mov(#2,#3){\sm*{\loop\ifnum\m*<\l*
\a*\m*\b* \advance\a*\d* \divide\a*\N* \multiply\a*\m*
\a**\m*\b** \advance\a**\d** \divide\a**\N* \multiply\a**\m*
\rmov*(\a*,\a**){\unhcopy\spl*}\advance\m*\*one\repeat}}}

\catcode`\*=12

\newcount\n@ast

\def\n@ast@#1{\n@ast0\relax\get@ast@#1\end}
\def\get@ast@#1{\ifx#1\end\let\next\relax\else
\ifx#1*\advance\n@ast1\fi\let\next\get@ast@\fi\next}

\newif\if@up \newif\if@dwn
\def\up@down@#1{\@upfalse\@dwnfalse
\if#1u\@uptrue\fi\if#1U\@uptrue\fi\if#1+\@uptrue\fi
\if#1d\@dwntrue\fi\if#1D\@dwntrue\fi\if#1-\@dwntrue\fi}

\def\halfcirc#1(#2)[#3]{{\Lengthunit=#2\Lengthunit\up@down@{#3}\relax
\if@up\mov(0,.5){\@arc[-][-]\@arc[+][-]}\fi
\if@dwn\mov(0,-.5){\@arc[-][+]\@arc[+][+]}\fi
\def\lft{\mov(0,.5){\@arc[-][-]}\mov(0,-.5){\@arc[-][+]}}\relax
\def\rght{\mov(0,.5){\@arc[+][-]}\mov(0,-.5){\@arc[+][+]}}\relax
\if#3l\lft\fi\if#3L\lft\fi\if#3r\rght\fi\if#3R\rght\fi
\n@ast@{#1}\relax
\ifnum\n@ast>0\if@up\shade[+]\fi\if@dwn\shade[-]\fi\fi
\ifnum\n@ast>1\if@up\dshade[+]\fi\if@dwn\dshade[-]\fi\fi}}

\def\halfdashcirc(#1)[#2]{{\Lengthunit=#1\Lengthunit\up@down@{#2}\relax
\if@up\mov(0,.5){\dash@arc[-][-]\dash@arc[+][-]}\fi
\if@dwn\mov(0,-.5){\dash@arc[-][+]\dash@arc[+][+]}\fi
\def\lft{\mov(0,.5){\dash@arc[-][-]}\mov(0,-.5){\dash@arc[-][+]}}\relax
\def\rght{\mov(0,.5){\dash@arc[+][-]}\mov(0,-.5){\dash@arc[+][+]}}\relax
\if#2l\lft\fi\if#2L\lft\fi\if#2r\rght\fi\if#2R\rght\fi}}

\def\halfwavecirc(#1)[#2]{{\Lengthunit=#1\Lengthunit\up@down@{#2}\relax
\if@up\mov(0,.5){\wave@arc[-][-]\wave@arc[+][-]}\fi
\if@dwn\mov(0,-.5){\wave@arc[-][+]\wave@arc[+][+]}\fi
\def\lft{\mov(0,.5){\wave@arc[-][-]}\mov(0,-.5){\wave@arc[-][+]}}\relax
\def\rght{\mov(0,.5){\wave@arc[+][-]}\mov(0,-.5){\wave@arc[+][+]}}\relax
\if#2l\lft\fi\if#2L\lft\fi\if#2r\rght\fi\if#2R\rght\fi}}

\catcode`\*=11

\def\Circle#1(#2){\halfcirc#1(#2)[u]\halfcirc#1(#2)[d]\n@ast@{#1}\relax
\ifnum\n@ast>0\L*=\xscale\Lengthunit
\ifnum\angle**=0\clap{\vrule width#2\L* height.1pt}\else
\L*=#2\L*\L*=.5\L*\special{em:linewidth .001pt}\relax
\rmov*(-\L*,0pt){\sm*}\rmov*(\L*,0pt){\sl*}\relax
\special{em:linewidth \the\linwid*}\fi\fi}

\catcode`\*=12

\def\wavecirc(#1){\halfwavecirc(#1)[u]\halfwavecirc(#1)[d]}
\def\dashcirc(#1){\halfdashcirc(#1)[u]\halfdashcirc(#1)[d]}

\def\xscale{1}

\def\yscale{1}

\def\Ellipse#1(#2)[#3,#4]{\def\xscale{#3}\def\yscale{#4}\relax
\Circle#1(#2)\def\xscale{1}\def\yscale{1}}

\def\dashEllipse(#1)[#2,#3]{\def\xscale{#2}\def\yscale{#3}\relax
\dashcirc(#1)\def\xscale{1}\def\yscale{1}}

\def\waveEllipse(#1)[#2,#3]{\def\xscale{#2}\def\yscale{#3}\relax
\wavecirc(#1)\def\xscale{1}\def\yscale{1}}

\def\halfEllipse#1(#2)[#3][#4,#5]{\def\xscale{#4}\def\yscale{#5}\relax
\halfcirc#1(#2)[#3]\def\xscale{1}\def\yscale{1}}

\def\halfdashEllipse(#1)[#2][#3,#4]{\def\xscale{#3}\def\yscale{#4}\relax
\halfdashcirc(#1)[#2]\def\xscale{1}\def\yscale{1}}

\def\halfwaveEllipse(#1)[#2][#3,#4]{\def\xscale{#3}\def\yscale{#4}\relax
\halfwavecirc(#1)[#2]\def\xscale{1}\def\yscale{1}}

\catcode`\@=\the\CatcodeOfAtSign

\title{On the effective potential for Horava-Lifshitz-like theories}

\author{C. F. Farias}

\affiliation{Departamento de F\'{\i}sica, Universidade Federal da 
Para\'{\i}ba\\
Caixa Postal 5008, 58051-970, Jo\~ao Pessoa, Para\'{\i}ba, Brazil}
\email{cffarias,jroberto,petrov@fisica.ufpb.br}

\author{M. Gomes}
\affiliation{Instituto de F\'\i sica, Universidade de S\~ao Paulo\\
Caixa Postal 66318, 05315-970, S\~ao Paulo, SP, Brazil}
\email{mgomes,ajsilva@fma.if.usp.br}

\author{J. R. Nascimento}

\affiliation{Departamento de F\'{\i}sica, Universidade Federal da
Para\'{\i}ba\\
Caixa Postal 5008, 58051-970, Jo\~ao Pessoa, Para\'{\i}ba, Brazil}
\email{cffarias,jroberto,petrov@fisica.ufpb.br}

\author{A. Yu. Petrov}

\affiliation{Departamento de F\'{\i}sica, Universidade Federal da 
Para\'{\i}ba\\
Caixa Postal 5008, 58051-970, Jo\~ao Pessoa, Para\'{\i}ba, Brazil}
\email{cffarias,jroberto,petrov@fisica.ufpb.br}

\author{A. J. da Silva}
\affiliation{Instituto de F\'\i sica, Universidade de S\~ao Paulo\\
Caixa Postal 66318, 05315-970, S\~ao Paulo, SP, Brazil}
\email{mgomes,ajsilva@fma.if.usp.br}

\begin{abstract}
We study the one-loop effective potential for some
Horava-Lifshitz-like theories.
\end{abstract}
\maketitle

The Horava-Lifshitz (HL) approach \cite{Hor} has recently acquired a
great scientific attention. This approach is characterized by an
essential asymmetry between space and time coordinates (space-time
anisotropy): the equations of motion of the theory are invariant under
the rescaling $x^i\to bx^i$, $t\to b^zt$, 
where $z$, the critical exponent, is a number characterizing its
ultraviolet behaviour. The main reason for it is that for the HL-like
reformulation of the known field theory models with a nontrivial
critical exponent $z>1$ leads to an improvement of the
renormalization of these models. In particular, the
four-dimensional gravity becomes renormalizable at $z=3$. 

Different
issues related to the HL gravity, including its cosmological  
aspects \cite{HorCos}, exact solutions \cite{Lu}, black holes \cite{BH} 
were considered in a number of papers. At the same time, the  study of 
the impacts of the HL extension to
other field theories is a very interesting problem. Some aspects of
the HL generalizations for the gauge field theories were presented in
\cite{ed}. Renormalizability of the scalar field theory models with
space-time anisotropy has been discussed in details in
\cite{Anselmi}. The four-fermion HL-like theory has been studied in
\cite{ff}. The Casimir effect for the HL-like scalar field theory has
been considered in \cite{ourcas}. In \cite{cpn}, the HL modifications
of the $CP^{N-1}$ were studied. The possibility of restoration of the
Lorentz symmetry in the theories with the space-time anisotropy is
discussed in  \cite{Gomes,ed}.

 It is well known that the effective potential is a key object in the
 quantum field theory useful for studying  many of its aspects. Some
 interesting results for  the HL-like theories have been obtained in
 the papers \cite{Eune,Liou} where the effective potential for the
 $\phi^4$  and the Liouville-Lifshitz theories have been studied. Also, some interesting results 
 for the effective potential in scalar field theories with certain values of the  critical exponent, have been obtained in \cite{Farakos}. In
 this paper, we intend to study the effective potential for a more
 generic class of theories including an arbitrary interaction of the
 scalar field with other fields. In the sequel, we will treat three
 cases, namely, a pure scalar model, a gauge model and a Yukawa model.

\paragraph{Scalar model.} We start with the straightforward HL
generalization 
of the usual scalar model:
\bea
S=\int dt d^dx
(\frac{1}{2}\dot{\phi}^2-\frac{1}{2}(-1)^z\phi\Delta^z\phi-
V(\phi)).
\eea
The renormalizability of such a model has been discussed in
\cite{Anselmi}. In general, renormalizability of such models requires
a polynomial form of the potential, however, for simplicity we
restrict ourselves to the form $V(\phi)=\lambda\phi^n$. 
Here our aim is the study of its effective
potential. To proceed with it, we, as usual, make the replacement
$\phi\to\Phi+\phi$, where $\Phi$ is a background field, and $\phi$ is
a quantum one. For the one-loop calculations, it is sufficient to keep
only the terms of the second order in the quantum field 
$\phi$:
\bea
S_2=-\frac{1}{2}\int dt d^dx\phi(\pa^2_0+(-1)^z\Delta^z+
V^{\prime\prime}(\Phi))\phi.
\eea
Following the standard procedure, the one-loop effective action can
be cast 
as
\bea
\label{g1l}
\Gamma^{(1)}=\frac{i}{2}{\rm Tr}\ln(\pa^2_0+(-1)^z\Delta^z+
V^{\prime\prime}(\Phi)).
\eea
The corresponding effective potential $U(\Phi)$ can be read off from the
expression
$$
\Gamma^{(1)}|_{\Phi=const}=-\int dt d^dx U^{(1)}(\Phi).
$$
To calculate $U(\Phi)$, we must carry out the Fourier
transform of (\ref{g1l}). After the 
Wick rotation, we arrive at
\bea
U^{(1)}=\frac{1}{2}\int\frac{dk_0d^dk}{(2\pi)^{d+1}}\ln(k^2_0+
\vec{k}^{2z}+V^{\prime\prime}(\Phi)).
\eea
First, we calculate the integral over $k_0$.  We use 
\bea
\frac{d}{d(A^2)}\int dk_0\ln(k^2_0+A^2)=\int\frac{dk_0}{k^2_0+A^2}=
\frac{\pi}{\sqrt{A^2}},
\eea
so that, neglecting an irrelevant field-independent constant, we get
\bea
U^{(1)}=\int\frac{d^dk}{(2\pi)^d}\sqrt{\vec{k}^{2z}+
V^{\prime\prime}(\Phi)}.
\eea
Then, we use the identity
\bea
\sqrt{B}=-\frac{1}{2\sqrt{\pi}}\int_{0}^{\infty} d\alpha \alpha^{-3/2} 
e^{-\alpha B}.
\eea
 Thus, 
\bea
U^{(1)}=-\frac{1}{2\sqrt{\pi}}\int d\alpha \alpha^{-3/2}
\int\frac{d^dk}{(2\pi)^d}e^{-\alpha(\vec{k}^{2z}+V^{\prime\prime}(\Phi))}.
\eea
In spherical coordinates and after the change of variables $k^z\to u$,
so, 
$k^{2z}=u^2$, $k=u^{1/z}$, and $dk=\frac{1}{z}du u^{1/z-1}$, we get 
\bea
U^{(1)}=-\frac{1}{2\sqrt{\pi}}\frac{1}{(2\pi)^d}\frac{1}{z}
\frac{2\pi^{d/2}}{\Gamma(d/2)}\int_0^{\infty} d\alpha \alpha^{-3/2}
\int_0^{\infty}du u^{\frac{d-z}{z}}e^{-\alpha(u^2+V^{\prime\prime}(\Phi))}.
\eea
 After integration we arrive at
\bea
U^{(1)}=-\frac{1}{2\sqrt{\pi}}\frac{1}{(2\pi)^d}\frac{1}{z}
\frac{\pi^{d/2}}{\Gamma(d/2)}\Gamma(\frac{d}{2z})\Gamma(-\frac{1}{2}-
\frac{d}{2z})
(V^{\prime\prime}(\Phi))^{1/2+d/(2z)}.
\eea
It is clear that this one-loop effective potential diverges if we have
$\frac{1}{2}(1+\frac{d}{z})=N$, where $N$ is a non-negative integer
number, in particular, for $z=2$, it diverges only at $d=2,6,10,..$. For
example, for $V(\Phi)\propto\Phi^{10}$, with $d=3$ and $z=2$, the
Green functions have a superficial degree of divergence
$\omega=5-\frac{E}{2}$, with $E$ is a number of legs. For the one-loop
renormalizability the model requires a counterterm $\Phi^8$. However,
the explicit calculation shows that such a correction is one-loop 
finite within the dimensional regularization. For $d=3$ that
expression is, as it
is well known, quadratically divergent for $z=1$  and
linearly divergent if $z=3$; otherwise it is finite.

\paragraph{Gauge fields.} Now, let us introduce gauge fields. For the
sake of concreteness, we restrict ourselves to the case $z=2$. In this
case, the Lagrangian of the scalar QED  is
\bea
\label{lasqed}
L=\frac{1}{2}F_{0i}F_{0i}+\frac{1}{4}F_{ij}\Delta F_{ij}-D_0\phi
(D_0\phi)^*+
D_iD_j\phi(D_iD_j\phi)^*-m^4\phi\phi^*,
\eea
where $D_0=\pa_0-ieA_0$, $D_i=\pa_i-ieA_i$ is a gauge covariant
derivative, with the corresponding gauge transformations: $\phi\to
e^{ie\xi}\phi$, $\phi^*\to e^{-ie\xi}\phi^*$, $A_0\to
A_0+\pa_0\xi$, $A_i\to A_i+\pa_i\xi$. To keep track only from the
gauge-matter 
interaction, we suggest that there is no self-coupling of the matter field.

The propagator for the scalar field has the simplest form
\bea
<\phi\phi^*>=\frac{i}{k^2_0-\vec{k}^4-m^4}.
\eea
As for the propagator of the gauge field, the situation is more
complicated. Indeed, to find this propagator, we must add to the free
Lagrangian of the gauge field
\bea
\label{free}
L_2=\frac{1}{2}F_{0i}F_{0i}+\frac{1}{4}F_{ij}\Delta
F_{ij}=\frac{1}{2}\pa_iA_0\pa_iA_0-\pa_0A_0\pa_iA_i+\frac{1}{2}\pa_0A_i\pa_0A_i
+\frac{1}{4}F_{ij}
\Delta F_{ij}.
\eea
the gauge-fixing term. However, since the $L_2$ contains a mixed term
involving both $A_0$ and $A_i$ (which have distinct behaviours), it
would be 
good if the gauge-fixing term could allow for the separation of these fields. 

It turns out to be that the appropriate gauge-fixing term is nonlocal:
\bea
L_{gf}=\frac{1}{2}(\frac{1}{\sqrt{\Delta}}\pa_0A_0+\sqrt{\Delta}\pa_iA_i)^2=
\frac{1}{2}(\pa_0A_0\frac{1}{\Delta}\pa_0A_0+2\pa_0A_0\pa_iA_i+\pa_jA_j\Delta
\pa_iA_i).
\eea
This gauge-fixing term can be treated as the analogue of the Feynman gauge.
Adding this gauge-fixing term to the $L_g$, we arrive at the following
complete Lagrangian:
\bea
L_c=L_2+L_{gf}=-\frac{1}{2}A_0\frac{\pa^2_0+\Delta^2}{\Delta}A_0-\frac{1}{2}
A_i(\pa^2_0+\Delta^2)A_i.
\eea
The nonlocality of this Lagrangian, however, does not give any danger
for  calculations. Indeed, the propagators have a reasonable form:
\bea
\label{props}
<A_0A_0>&=&\frac{i\vec{k}^2}{k^2_0-\vec{k}^4};\nonumber\\
<A_iA_j>&=&-\frac{i\delta_{ij}}{k^2_0-\vec{k}^4}.
\eea
As can be checked, the model is then renormalizable for $d\leq 4$.

To calculate the effective potential, we must take into account that
it depends only on the matter fields, thus, we treat the gauge field
as a pure quantum field. Also, we must take into account that, within
the one-loop approximation, only the vertices associated to two
quantum fields give nontrivial contributions to the effective
potential. Let us denote the background fields by $\Phi$ and
$\Phi^*$. It is easy to see that the only relevant  vertices are
\bea
&&-e^2A_0A_0\Phi\Phi^*;\quad\, ieA_0(\Phi^*\pa_0\phi-\Phi\pa_0\phi^*),
\nonumber\\
&&-ie(\pa_iA_j)[\Phi\pa_i\pa_j\phi^*-\Phi^*\pa_i\pa_j\phi],\quad\,  
e^2(\pa_iA_j)(\pa_iA_j)\Phi\Phi^*.
\eea
To simplify the calculations, it is convenient to move within these
vertices all derivatives to act on the gauge fields. So, these
vertices take  the form:
\bea
\label{racvert}
&&-e^2A_0A_0\Phi\Phi^*;\quad\, -ie(\Phi^*\phi-\Phi\phi^*)\pa_0 A_0,\nonumber\\
&&-ie[\Phi\phi^*-\Phi^*\phi](\pa_j\Delta A_j),\quad\,  
-e^2A_j\Delta A_j\Phi\Phi^*.
\eea

To fix the quantum corrections at the one-loop order, we must consider
two 
types of
contributions. In the first of them, all diagrams involve only the
gauge field  propagators in the internal lines:

\vspace*{2mm}

\hspace{2.0cm}
\Lengthunit=1.2cm
\Linewidth{.5pt}
\GRAPH(hsize=4){\mov(.5,0){\wavecirc(1)}\mov(.5,.5){\lin(-.5,.5)\lin(.5,.5)}
\mov(3,0){\mov(.5,0){\wavecirc(1)}\mov(.5,.5){\lin(-.5,.5)\lin(.5,.5)}
\mov(.5,-.5){\lin(-.5,-.5)\lin(.5,-.5)}}
\mov(6,0){\mov(.5,0){\wavecirc(1)}\mov(.5,.5){\lin(-.5,.5)\lin(.5,.5)}
\mov(.5,-.5){\lin(-.5,-.5)\lin(.5,-.5)}\mov(.9,0){\lin(.5,-.5)\lin(.5,.5)}}
\ind(80,0){\ldots}
}

\vspace*{2mm}

The total result from this sector is a sum of two contributions to the
effective potential -- the
first one,
$U_a$ is given by sum of loops of $<A_0A_0>$ propagators, and the
second one, $U_b$ -- of $<A_iA_j>$ propagators:
\bea
U_a&=&-\sum\limits_{n=1}^{\infty}\frac{1}{n}\int
\frac{d^dkdk_0}{(2\pi)^{d+1}}(e^2\Phi\Phi^*)^n
\Big(\frac{\vec{k}^2}{k^2_0-\vec{k}^4}
\Big)^n;\nonumber\\
U_b&=&-\sum\limits_{n=1}^{\infty}\frac{d}{n}\int
\frac{d^dkdk_0}{(2\pi)^{d+1}}(e^2\Phi\Phi^*)^n
\Big(\frac{\vec{k}^2}{k^2_0-\vec{k}^4}
\Big)^n.
\eea

The second type of diagrams involves the triple vertices as well. 
We should first introduce a "dressed" propagator

\hspace{4.0cm}
\Lengthunit=1.2cm
\GRAPH(hsize=4){\Linewidth{1.5pt}\wavelin(1,0)\ind(12,0){=}\Linewidth{0.5pt}
  \mov(1.5,0){\wavelin(1,0)}\ind(28,0){+}\mov(3,0){\wavelin(2,0)
\mov(1,0){\lin(-.5,.5)\lin(.5,.5)}}\ind(53,0){+}\ind(58,0){\ldots}
}

In this propagator, the summation over all quartic vertices is
performed. As a  result, these "dressed" propagators are equal to
\bea
\label{dress}
<A_0A_0>_D&=&<A_0A_0>\sum\limits_{n=0}^{\infty}[ie^2\Phi\Phi^*<A_0A_0>]^n= 
\frac{i\vec{k}^2}{k^2_0-\vec{k}^4-e^2\vec{k}^2\Phi\Phi^*};\nonumber\\
<A_iA_j>_D&=&-\frac{i\delta_{ij}}{k^2_0-\vec{k}^4}\sum\limits_{n=0}^{\infty}
[e^2\frac{\vec{k}^2}{k^2_0-\vec{k}^4}\Phi\Phi^*]^n=
-\frac{i\delta_{ij}}{k^2_0-\vec{k}^4-e^2\vec{k}^2\Phi\Phi^*}.
\eea

To proceed, we follow the methodology developed in \cite{GR} and other
papers. It is based on the summation over diagrams representing
themselves as cycles of all possible number of links. 
Such diagrams look like

\vspace*{2mm}

\hspace{6.0cm}
\Lengthunit=1cm
\GRAPH(hsize=2){\Linewidth{1.6pt}\halfwavecirc(2)[u]\Linewidth{.4pt}
\mov(-.1,0){\halfcirc(2)[d]\mov(-1,0){\lin(-.5,0)}\mov(1.05,0){\lin(.5,0)}}
}
\GRAPH(hsize=2){\mov(0,1){
\mov(.3,0){\Linewidth{1.6pt}\wavelin(2,0)\mov(0.05,-2){\wavelin(2,0)}}
\Linewidth{.4pt}
\lin(-.5,.5)\lin(0,-2)\mov(0,-2){\lin(-.5,-.5)}
\mov(1.9,0){\lin(.5,.5)\lin(0,-2)}\mov(1.9,-2){\lin(.5,-.5)}\ind(30,-10){\ldots}
}
}

\vspace*{2mm}

Now, it is time to take into account the derivatives in the triple
vertices. Using the "rationalized" form of the vertices
(\ref{racvert}), we can find that effectively one must  consider the objects
\bea
G_1&=&<\pa_0A_0(t_1,\vec{x}_1)\pa_0A_0(t_2,\vec{x}_2)>_D;\nonumber\\
G_2&=&<\pa_i\Delta A_i(t_1,\vec{x}_1)\pa_j \Delta A_j(t_2,\vec{x}_2)>_D,
\eea
whose Fourier transforms are
\bea
\label{efpr}
G_1(k)&=&\frac{ik^2_0\vec{k}^2}{k^2_0-\vec{k}^4-e^2\vec{k}^2\Phi\Phi^*};
\nonumber\\
G_2(k)&=&-\frac{i\vec{k}^6}{k^2_0-\vec{k}^4-e^2\vec{k}^2\Phi\Phi^*}.
\eea
Here we took into account that the derivatives affect different
arguments of  the propagator which changes the sign with respect to 
(\ref{dress}).
Then, we can take into account that the effective propagators $G_1$
and $G_2$ enter the diagrams above on the same base, thus, the total
contribution must be symmetric under replacement $G_1\leftrightarrow
G_2$. Thus, the total  contribution from these graphs is
\bea
\label{sumgc1}
U_c&=&-\sum\limits_{n=1}^{\infty}\frac{1}{n}\int
\frac{d^dkdk_0}{(2\pi)^{d+1}}(-e^2\Phi\Phi^*)^n
\Big((G_1+G_2)<\phi\phi^*>\Big)^n,
\eea
which yields
\bea
\label{gc}
U_c&=&-\sum\limits_{n=1}^{\infty}\frac{1}{n}\int
\frac{d^dkdk_0}{(2\pi)^{d+1}}(e^2\Phi\Phi^*)^n
\Big(\frac{(k^2_0-\vec{k}^4)\vec{k}^2}{k^2_0-\vec{k}^4-e^2\vec{k}^2\Phi\Phi^*}
\frac{1}{k^2_0-\vec{k}^4-m^4}\Big)^n.
\eea
It remains to process all these expressions
$U_a,U_b$ and $U_c$. To do  it, we use the identity 
$\sum\limits_{n=1}^{\infty}\frac{a^n}{n}=-\ln(1-a)$ and carry out the
Wick 
rotation, thus,
\bea
\label{corrs}
U_a&=&i\int\frac{d^dkdk_0}{(2\pi)^{d+1}}\ln[1+
\frac{e^2\Phi\Phi^*\vec{k}^2}{k^2_0+\vec{k}^4}];\nonumber\\
U_b&=&id\int\frac{d^dkdk_0}{(2\pi)^{d+1}}\ln[1+
\frac{e^2\Phi\Phi^*\vec{k}^2}{k^2_0+\vec{k}^4}];\nonumber\\
U_c&=&i\int\frac{d^dkdk_0}{(2\pi)^{d+1}}\ln[1-
\frac{e^2\Phi\Phi^*(k^2_0+\vec{k}^4)\vec{k}^2}{k^2_0+
\vec{k}^4+e^2\vec{k}^2\Phi\Phi^*}\frac{1}{k^2_0+\vec{k}^4+m^4}].
\eea
In the case $m=0$, $U_c$ simplifies radically, and we have
\bea
U_c&=&-i\int\frac{d^dkdk_0}{(2\pi)^{d+1}}\ln[1+\frac{e^2\Phi\Phi^*
\vec{k}^2}{k^2_0+\vec{k}^4}],
\eea
which completely cancels $U_a$. So, in this case we end just with the
following contribution to the  effective potential:
\bea
\label{int1}
U^{(1)}&=&id\int\frac{d^dkdk_0}{(2\pi)^{d+1}}\ln[1+ 
\frac{e^2\Phi\Phi^*\vec{k}^2}{k^2_0+\vec{k}^4}].
\eea
Adding and subtracting the constant
$id\int\frac{d^dkdk_0}{(2\pi)^{d+1}}\ln[1+\frac{\vec{k}^4}{k^2_0}]$,
we find that the effective potential, up to an additive  constant, looks like
\bea
U^{(1)}&=&id\int\frac{d^dkdk_{0E}}{(2\pi)^{d+1}}\ln[1+
\frac{\vec{k}^4+e^2\Phi\Phi^*\vec{k}^2}{k^2_0}].
\eea
Then, we use the integral $\int_0^{\infty}dk_0\ln(k^2_0+A^2)=\pi\sqrt{A^2}$, so,
\bea
\label{epqed}
U^{(1)}&=&id\int\frac{d^dk}{2(2\pi)^d}\sqrt{\vec{k}^2(\vec{k}^2+
e^2\Phi\Phi^*)}=idI.
\eea
Following the same steps as before, in the case of the scalar model,
we arrive at
\bea
\label{intf}
I=\frac{\pi^{d/2}}{2(2\pi)^d}(e^2\Phi\Phi^*)^{d/2+1}\frac{\Gamma(-1-
\frac{d}{2})\Gamma(\frac{d}{2}+\frac{1}{2})}{\Gamma(\frac{d}{2})
\Gamma(-\frac{1}{2})}.
\eea
We see that for odd spatial dimension $d$, this expression is
finite, while for even $d$ it diverges, and, we would need
to add the corresponding counterterms (in particular, for $d=2$, one
will need 
the quartic interaction to achieve a multiplicative renormalizability).

For completeness, we note that sometimes, the Coulomb gauge
$\pa_iA_i=0$ maybe convenient. It is considered in Appendix.

\paragraph{Yukawa theory.} Then, let us formulate the Yukawa
theory. It is natural to consider now the $z=2$ version of the spinor
field theory, so, the $(d+1)$-dimensional Lagrangian for the theory
looks  like
\bea
\label{yukawa}
L=\bar{\psi}(i\gamma^0\pa_0+\Delta-m^2-h\Phi)\psi.
\eea
To keep track only from the Yukawa coupling, we treat the scalar field
as a purely external one. The generalization of this study for the
case of the self-interacting scalar field is  straightforward.
The one-loop effective potential corresponding to this Lagrangian, looks like
\bea
\Gamma^{(1)}=i{\rm Tr}\ln(i\gamma^0\pa_0+\Delta-m^2-h\Phi).
\eea
We can present this expression as
\bea
\Gamma^{(1)}=i{\rm Tr}\ln(i\gamma^0\pa_0)+
i{\rm Tr}\ln(1-i\frac{(\Delta-m^2-h\Phi)\gamma^0\pa_0}{\pa^2_0}).
\eea
Disregarding an irrelevant additive constant, expanding the logarithm
in power series, calculating the matrix trace and doing the sum, we arrive at
\bea
\Gamma^{(1)}=i\frac{\delta}{2}{\rm Tr}
\ln\Big[1-\frac{(\Delta-m^2-h\Phi)^2}{\pa^2_0}
\Big].
\eea
Here $\delta$ is a dimension of the Dirac matrices in the
corresponding representation.
After Fourier transform by the rule $i\pa_{0,i}\to k_{0,i}$, this expression 
yields the following effective potential
\bea
U^{(1)}=-i\frac{\delta}{2}\int\frac{d^dkdk_0}{(2\pi)^{d+1}}
\ln\Big[\frac{k^2_0-(\vec{k}^2+m^2+h\Phi)^2}{k^2_0}
\Big].
\eea
Doing the Wick rotation and integrating over $k_0$, we arrive at
\bea
U^{(1)}=-\frac{\delta}{2}\int\frac{d^dk}{(2\pi)^d}(\vec{k}^2+m^2+
h\Phi)^{1+\epsilon}.
\eea
This integral, for any positive $d$, vanishes within the dimensional  
regularization being proportional to $\frac{1}{\Gamma(-1-\epsilon)}$
which is zero as $\epsilon\to 0$. 

An observation is in order: if we
consider a model composed of the Lagrangian (\ref{lasqed}) plus an
extension of (\ref{yukawa}) in which the fermions are also minimally
coupled to the electromagnetic field, up to the one-loop  order, no
additional contribution to the effective potential given by the
expressions (\ref{epqed},\ref{intf}) arises.

We studied the effective action for some scalar HL-like theories: first, the self-coupled
scalar model whose one-loop effective potential was found for 
arbitrary values of the space dimension, critical exponent and
coupling; second, the scalar QED, whose effective potential was
successfully obtained in the $z=2$ case, and third,  the Yukawa theory, where
the one-loop effective potential was shown to vanish. In principle, we
can also introduce the coupling between the gauge and spinor
fields. Nevertheless, these additional interactions  will start to contribute to the effective
potential only at the two-loop order.
We found that the
methodology for calculating the effective potential does not
essentially differ from that in the usual, Lorentz invariant field theories.

\centerline{\bf Appendix}

Let us briefly describe the difference of
the results for the  case of gauge fields in the Coulomb gauge.
After imposing this gauge, the "mixed" term immediately vanishes in
the action (\ref{free}), but there is no modification of the quadratic
term in $A_0$, so, the propagator $<A_0A_0>$  in this case differs
from that  one in (\ref{props}) being equal to
\bea
<A_0A_0>=-\frac{i}{\vec{k}^2},
\eea
whereas the propagator $<A_iA_j>$ stays the same as in (\ref{props}). As
a result, the contribution $U_b$ from (\ref{corrs}) stays
unchanged, while for $U_a$ now we  have
\bea
\label{ga}
U_a&=&i\int\frac{d^dkdk_{0E}}{(2\pi)^{d+1}}\ln[1+
\frac{e^2\Phi\Phi^*}{\vec{k}^2}].
\eea 
However, the result for $\Gamma_c$ in the Coulomb gauge 
is much more complicated. Let us
consider it in details. 
The "effective propagator" $G_1$ introduced in (\ref{efpr}) in the
case of the Coulomb gauge takes the  form
\bea
G_1=-\frac{ik^2_0}{\vec{k}^4+e^2\vec{k}^2\Phi\Phi^*},
\eea
while the $G_2$ does not suffer any modification. After doing the sum
indicated in (\ref{sumgc1}) and some simple transformations, we have
\bea
U_c&=&i\int\frac{d^dkdk_0}{(2\pi)^{d+1}}
\Big[\ln\big[\frac{[k^2_0\vec{k}^2+(\vec{k}^2+e^2\Phi\Phi^*)(\vec{k}^4+m^4)]
(k^2_0+\vec{k}^4+e^2\vec{k}^2\Phi\Phi^*)}{k^4_0\vec{k}^2}+\nonumber\\ &+&
\frac{e^2\Phi\Phi^*\vec{k}^6(\vec{k}^2+e^2\Phi\Phi^*)
}{k^4_0\vec{k}^2}\big]-\ln\frac{k^2_0+\vec{k}^4+
e^2\vec{k}^2\Phi\Phi^*}{k^2_0}-\ln\frac{\vec{k}^2+e^2\Phi\Phi^*}{
\vec{k}^2}+\ln\frac{k^2_0}{\vec{k}^4}\Big].
\eea
It is easy to see that the second term in the r.h.s. of this
expression differs from (\ref{int1}) only by a constant factor, $-d$, 
multiplying the
last one. The third term exactly cancels with $U_a$ (\ref{ga}),
and the last term is a pure irrelevant constant, since it does not depend on the background fields. Thus,
the complete effective potential is
\bea
U&=&U_a+U_b+U_c=
-i\frac{\pi^{d/2}}{2(2\pi)^d}(e^2\Phi\Phi^*)^{d/2+1}\frac{(1-d)\Gamma(-1-
\frac{d}{2})\Gamma(\frac{d}{2}+\frac{1}{2})}{\Gamma(\frac{d}{2})
\Gamma(-\frac{1}{2})}+\nonumber\\
&+&
i\int\frac{d^dkdk_0}{(2\pi)^{d+1}}
\ln\big[\frac{[k^2_0\vec{k}^2+(\vec{k}^2+e^2\Phi\Phi^*)(\vec{k}^4+m^4)]
(k^2_{0E}+\vec{k}^4+e^2\vec{k}^2\Phi\Phi^*)}{k^4_0\vec{k}^2}+\nonumber\\ &+&
\frac{e^2\Phi\Phi^*\vec{k}^6(\vec{k}^2+e^2\Phi\Phi^*)
}{k^4_0\vec{k}^2}\big].
\eea
Notice that the first term in this expression is very similar to the
result (\ref{ga}). The above expression differs from the results in 
(\ref{corrs}--\ref{intf}) just because the contribution involving \\
$<A_0A_0>$ propagators was cancelled. 
Unfortunately, the last term is highly cumbersome.

 The discrepancy between the results for the effective
potential in the  gauges we considered is expected because it is a gauge dependent
quantity.

{\bf Acknowledgements.} This work was partially supported by Conselho
Nacional de Desenvolvimento Cient\'{\i}fico e Tecnol\'{o}gico (CNPq)
and Funda\c{c}\~{a}o de Amparo \`{a} Pesquisa do Estado de S\~{a}o
Paulo (FAPESP), and CNPq/PRONEX/FAPESQ. The work by A. Yu. P. has been supported by the CNPq, project 303461/2009-8.

\end{document}